\documentclass{article}
\usepackage{hiph-preprint}
\usepackage{graphicx}
\usepackage{amssymb}

\newcommand{\Lam}{\ensuremath{\Lambda}}
\newcommand{\ALam}{\ensuremath{\bar{\Lambda}}}

\volnumber{00} \issuenumber{0} \edyear{2006}                             
\frompage{000} \topage{000}                                              
\recrevdate{8 February 2006}                                              
\def\rightmark{Forward \Lam\ Production and Nuclear Stopping Power at RHIC}
\def\leftmark{F.~Simon}
 
\title{Forward $\Lambda$ Production and Nuclear Stopping Power in d+Au Collisions at RHIC} 
\authors{{F.~Simon$^{1,a}$ for the STAR Collaboration %
\index{Simon, F.} 
}\\[2.812mm]
{\normalsize
\hspace*{-8pt}$^{1}$ Massachusetts Institute of Technology, Cambridge, MA 02139, USA\\[0.2ex] 
}}
 
\abstract{Using the forward time projection chambers of STAR we measure the
centrality dependent \Lam\ and \ALam\ yields in d+Au collisions at 
$\sqrt{s_{NN}}$ = 200 GeV at forward and backward rapidities.
The contributions of different processes to particle production and
baryon transport are probed exploiting the inherent asymmetry of 
the d+Au system. While the d side appears to be dominated by multiple independent nucleon-nucleon collisions,
nuclear effects contribute significantly on the Au side. Using the constraint of baryon number 
conservation, the rapidity loss of baryons in the incoming deuteron can 
be estimated as a function of centrality. This is compared to a model and to similar measurements
in Au+Au, which gives insights into the nuclear stopping power at
relativistic energies.
}

 
\makeindex
\begin{document}
 
\maketitle

\section{Introduction}
\label{sec:intr} \setcounter{section}{1}\setcounter{equation}{0}

In asymmetric heavy ion collisions studies of particle production away from mid-rapidity are of particular interest since this allows to distinguish between effects stemming from the projectile and the target. While the deuteron side of the reaction is anticipated to be dominated by an overlap of multiple independent nucleon-nucleon collisions of the deuteron nucleons with gold nucleons, nuclear effects are expected to contribute significantly on the gold side of the interaction. By a comparison of the data to model calculations this can be tested. 

In addition, the collision of a light projectile with a heavy target can be used to study the nuclear stopping power, a quantity of fundamental interest in heavy ion physics \cite{Stopping}, since it is related to the energy lost by incoming baryons and thus available for the production of possible new states of matter. Substantial rapidity loss of baryons has been observed in central Au+Au collisions at the highest energy available at the Relativistic Heavy Ion Collider (RHIC) \cite{Brahms}. In d+Au collisions the baryon transport is expected to be influenced by the different mechanisms on the two sides of the collision, leading to appreciable differences between the target and the projectile sides of the reaction.

In the present paper, the production of \Lam\ hyperons and their anti-particles is studied in d+Au collisions at $\sqrt{s_{NN}}$ = 200 GeV at forward and backward rapidities ($y$ = $\pm$ 2.75). This is done using the radial-drift forward TPCs (FTPCs) \cite{FTPC} of the STAR experiment at RHIC, which measure charged hadrons in the pseudorapidity range 2.5 $<$ $\vert\eta\vert$ $<$ 4.0. The detector that sits at negative rapidity and thus intercepts the fragments of the Au after the reaction is referred to as FTPC-Au, the detector on the d side as FTPC-d. The \Lam\ particles are reconstructed using their decay \Lam\ $\rightarrow$ p$\pi$, via displaced decay vetices formed by oppositely charged tracks \cite{SQM}. The data sample of 10.4 million minimum bias events is subdivided into central (top 20\%), mid-central (20\%--40\%) and peripheral (40\%--100\%) events, based on the charged particle multiplicity at mid-rapidity.

\section{Particle Yields and Model Comparisons}

\begin{figure}[t]
\begin{minipage}[t]{0.49\linewidth}
\begin{center}
\includegraphics[width=\linewidth]{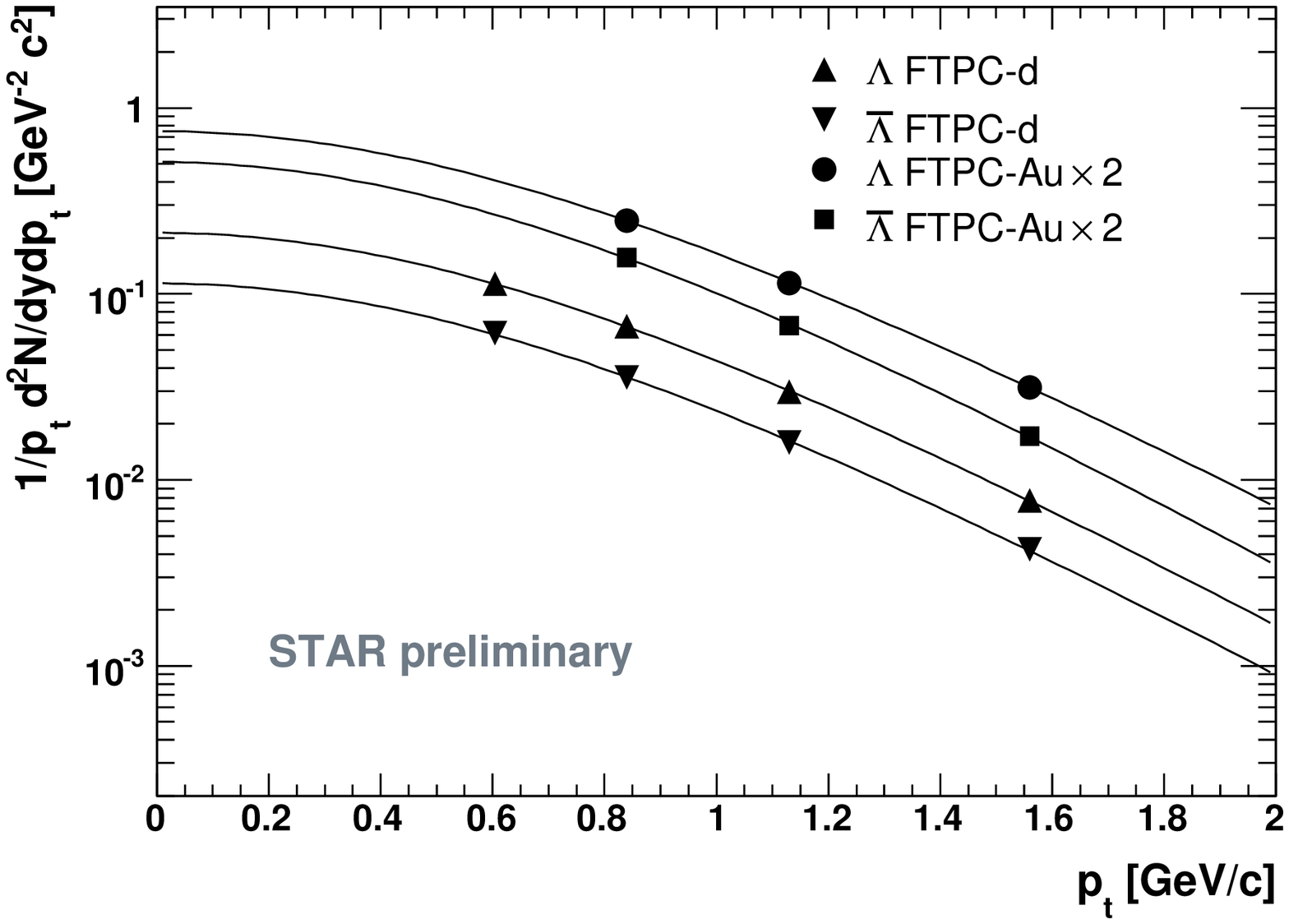} 
\end{center}
\caption{Transverse momentum spectra of \Lam\ and \ALam\ for d+Au minimum bias events on the d and the Au side of the collision. The yields are extracted with an $m_t$ exponential fit. Due to an incorrect treatment of electronics failure no reliable measurement at $p_t$ = 0.6 GeV/c could be obtained on the Au side.}
 \label{fig:1}
\end{minipage}%
\hspace{0.02\textwidth}
\begin{minipage}[t]{0.49\linewidth}
\begin{center}
\includegraphics[width=\linewidth]{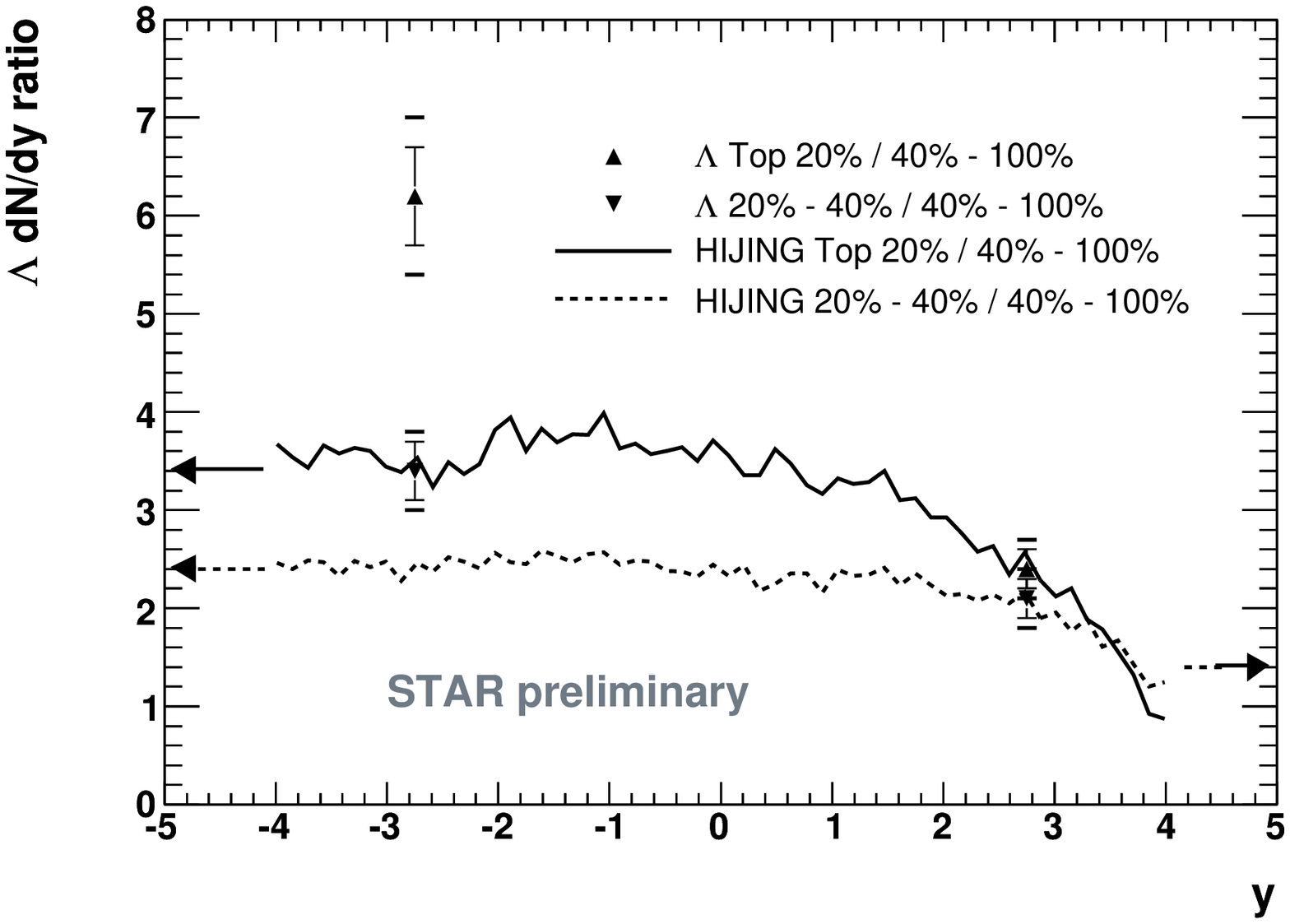} 
\end{center}
 \caption{Ratio of the \Lam\ yield in central and mid-central events to that in peripheral events. The lines show predictions by HIJING, the arrows indicate the values expected at extremely forward and backward rapidity in the case of a scaling with the number of deuteron and gold participants.}
\label{fig:2}
\end{minipage}
\end{figure}

Figure \ref{fig:1} shows the minimum bias transverse momentum spectra for \Lam\ and \ALam\ on both sides of the collision together with exponential fits in transverse mass $m_t$. The particle yields are obtained by integration of the fits. 

The evolution of the \Lam\ yield with collision centrality is shown in figure \ref{fig:2}. Here, the ratio of the yield in central to peripheral and in mid-central to peripheral d+Au events is plotted at forward and backward rapidity. The predictions from HIJING \cite{HIJING} are indicated by the lines, the expectations for very forward rapidities in the case of a scaling with the number of wounded d and Au nucleons are shown by arrows. Both the HIJING predictions and the expectations from wounded nucleon scaling agree well with the data on the deuteron side, while the increase of the particle yield on the Au side is significantly underpredicted. This indicates that nuclear effects, not incorporated in HIJING, play an important role on the Au side of the reaction.

\begin{figure}[t]
\begin{center}
\includegraphics[width=0.98\linewidth]{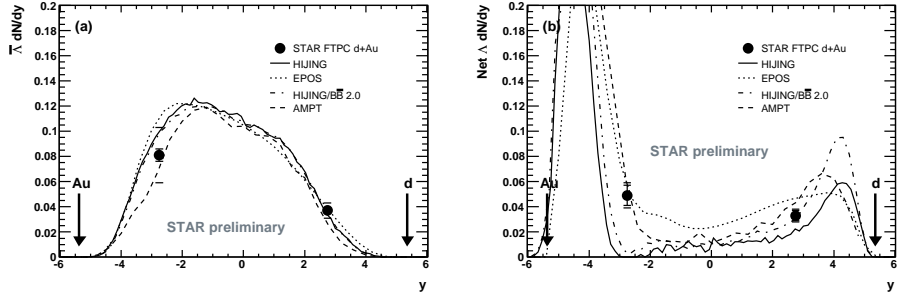}
\end{center}
\caption{Comparison of minimum bias yields with model calculations: a) \ALam\ dN/dy, b) net \Lam\ dN/dy. In both panels the target and projectile beam rapidities are indicated by arrows.}
\label{fig:3}
\end{figure}

By comparing the minimum bias particle yields to a variety of model calculations, the processes that drive the particle production and the transport of baryon number on both collision sides are investigated, since the models incorporate different mechanisms. HIJING is based on the superposition of individual nucleon-nucleon collisions, HIJING/B$\overline{\mbox{B}}$ \cite{HIJINGBBbar} provides increased baryon number transport via gluon junctions. AMPT \cite{AMPT} is a multi-phase model that has a hadronic transport stage in addition to HIJING-like initial processes. EPOS \cite{EPOS} includes nuclear effects via target and projectile remnants. Figure \ref{fig:3}a) shows the \ALam\ dN/dy together with the model prediction. On the d side there is excellent agreement between the data and all models, also on the Au side all models agree with the measurement within errors, pointing to a good understanding of \ALam-\Lam\ pair production in the collision. Figure \ref{fig:3}b) shows the net \Lam\ dN/dy together with model predictions. On the d side, HIJING/B$\overline{\mbox{B}}$ shows the best agreement, while on the Au side only AMPT and EPOS, which both incorporate nuclear effects, are able to describe the data. This suggests that nuclear effects have a sizable impact on baryon number transport on the Au side, while nuclear stopping on the d side is well described by multiple independent nucleon-nucleon collisions.

\section{Nuclear Stopping Power}

\begin{figure}[t]
\begin{minipage}[t]{0.49\linewidth}
\begin{center}
\includegraphics[width=\linewidth]{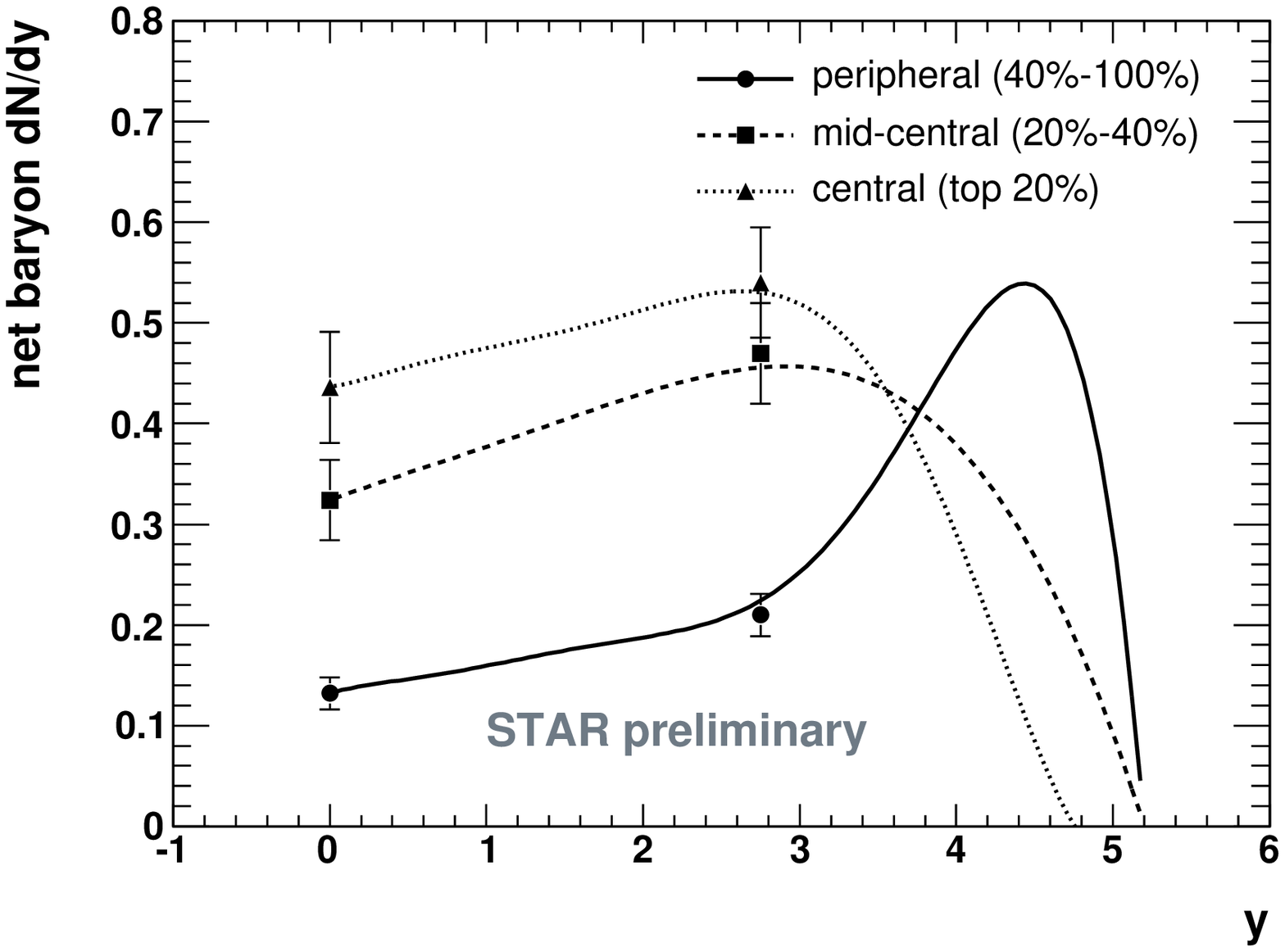} 
\end{center}
\caption{Net baryon dN/dy for three centralities on the d side, obtained from mid-rapidity TOF measurements and from the net \Lam\ results at forward rapidity. The lines show fits with 6$^{th}$ order polynomials with the integral of the number of deuteron particpants, giving an impression of the possible distribution of baryons in the high rapidity region not covered by the measurement.}
 \label{fig:4}
\end{minipage}
\hspace{0.02\textwidth}
\begin{minipage}[t]{0.49\linewidth}
\begin{center}
\includegraphics[width=\linewidth]{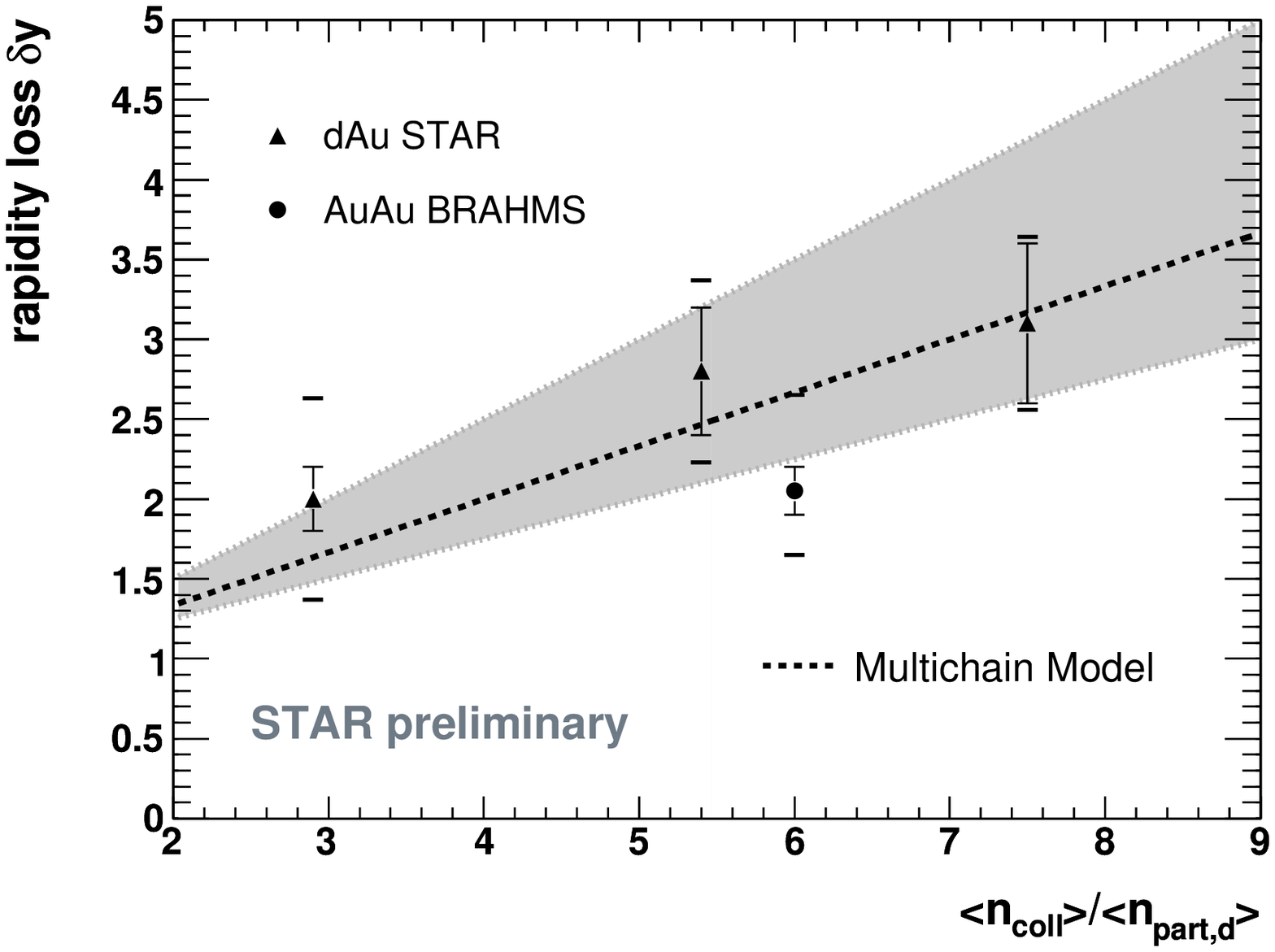} 
\end{center}
 \caption{Rapidity loss on the d side as a function of centrality, given by the number of binary collisions divided by the number of d participants. The error bars show the contributions from the net baryon yield estimate and the contribution from the extrapolation of the distribution at high rapidity. The results are compared to the multi-chain model \cite{MCM} (shaded) and measurements in central Au+Au collisions \cite{Brahms}.}
\label{fig:5}
\end{minipage}
\end{figure}

On the d side, a significant fraction of the total available baryon number, given by the number of d participants, is moved into the rapidity range accessible to measurements. It is thus possible to obtain an estimate of the mean rapidity loss of baryons from the d after their collision with the Au nucleus. To do this, total net baryon yields have to be estimated from the measured net \Lam\ yields. Simulations and mid-rapidity measurements yield the following correspondence: net baryons = ($10\pm 1$) $\times$ net \Lam. At mid-rapidity, proton measurements done with the STAR TOF detector \cite{TOF} are used to obtain the net baryon density. Figure \ref{fig:4} shows the estimated net baryon density at mid and forward rapidity. These are fitted with a 6$^{th}$ order polynomial with an integral set to the number of deuteron participants to give an impression of the possible distribution of baryons. This relies on the assumption that the baryon number in the deuteron is mostly confined to the deuteron side of the collision, and that the contribution from Au baryons is small.

By integration of the fits, the mean rapidity of the baryons after the collision is determined. The rapidity loss is obtained by $\delta y = y_{\Lambda, beam} - \left<y\right>$ and plotted in figure \ref{fig:5} with uncertainties stemming from the estimation of the net baryon yield and the extrapolation of the not measured region at high rapidity. The comparison to the rapidity loss measured in central Au+Au collisions suggests that a slightly higher rapidity loss per collision is reached in d+Au reactions. The good agreement of the d+Au data with the multi-chain model indicates that up to RHIC energies the rapidity loss in $p$(d)+A collisions is largely energy independent and depends primarily on the number of collisions in the reaction. 

\section{Conclusion}

The measured \Lam\ and \ALam\ yields at forward and backward rapidity in d+Au collisions at RHIC suggest that the d side of the reaction is well described by multiple individual nucleon-nucleon collisions, while the Au side is influenced significantly by nuclear effects. The nuclear stopping power, characterized by the rapidity loss of d baryons, increases linearly with collision centrality and is consistent with earlier data at lower energy.

\section*{Notes} 
\begin{notes}
\item[a] 
E-mail: fsimon@mit.edu
\end{notes}

\vfill\eject
\end{document}